\documentclass{PoS}
\usepackage{amsmath}
\usepackage{amssymb}
\usepackage{enumerate}
\usepackage{amsfonts}
\usepackage{epstopdf}
\usepackage[latin1]{inputenc}
\newcounter{saveeqn}

\title{Boson Stars and QCD Boson Stars}

\ShortTitle{Boson Stars and QCD Boson Stars}

\author{\speaker{Usha Kulshreshtha} \thanks{Invited Talk.}\\
Department of Physics, Kirori Mal College\\
University of Delhi, Delhi-110007, India.\\
        E-mail: \email{ushakulsh@gmail.com}} 
 
       \author{{Sanjeev Kumar} \\
Department of Physics\\
Pandit Neki Ram Sharma Government Post Graduate College \\ Rohtak - 124001, Haryana, India.\\
        E-mail: \email{sanjeev.kumar.ka@gmail.com}}

\author{{Daya Shankar Kulshreshtha} \\
Department of Physics and Astrophysics\\
University of Delhi, Delhi-110007, India.\\
        E-mail: \email{dskulsh@gmail.com}}

   \author{{Jutta Kunz} \\
Institute of Theoretical Physics\\
Carl von Ossietzky University of Oldenburg\\
D 26111 Oldenburg, Germany\\
        E-mail: \email{jutta.kunz@uni-oldenburg.de}}  
        
\abstract{In this talk, we present a review of our work on boson stars in a theory of a massless complex scalar field in the presence of a $U(1)$ gauge field and gravity. A sequence of bifurcation points obtained in the phase diagrams of the theory is presented and the plots of the mass $~M~$ versus charge $~Q~$ as well as plots of the mass per unit charge $~M/Q~$ versus the charge $~Q~$ of the boson stars are presented along with a discussion of the results. We also present some ideas on the possibilities of QCD boson stars.}

\FullConference{Light Cone 2019 - LC2019\\
	 September 16-20, 2019\\
		Ecole Polytechnique, Palaiseau, France}

\begin{document}

Charged  compact boson stars and boson shells represent localized self-gravitating solutions with sharp boundaries \cite{1}-\cite{9}. They could consist of any kind of particle that could be represented by scalar fields like the scalar axion, the Higgs boson, the quark-antiquark pair or diquarks or anti-diquarks etc. 
One could consider these theories with or without a cosmological constant $~\Lambda~$ which could take positive or negative values corresponding to the de-Sitter (dS) and Anti de-Sitter (AdS) spaces respectively \cite{1}-\cite{9}.

In the following we present a review of our work on a study of boson stars in a theory of a massless complex scalar field in the presence of a $~U(1)~$ gauge field and gravity. The  phase diagrams (PDs) obtained in our theory are presented. A sequence of bifurcation points (BPs) is seen to exist in the PDs of the theory. We also present plots of mass $~M~$ versus charge $~Q~$ as well as plots of the mass per unit charge $~M/Q~$ versus the charge $~Q~$ of the boson stars along with a discussion of the results. We also present some ideas on the possibilities of QCD boson stars \cite{10}.

For this we consider boson stars and boson shells in a theory of a complex scalar field $~\Phi~$ coupled to a $~U(1)~$ gauge field and gravity in the presence of a conical potential: $~V(|\Phi|)~$ (which is a function of $~|\Phi|~$) defined by: $~V(|\Phi|) := \lambda |\Phi|~$ (where $~\lambda~$ is a constant parameter). 
We study the  properties of the solutions of this theory and determine their domains of existence for some specific values of the parameters of the theory. The theory could be  studied with or without a mass term for the complex scalar field \cite{3}-\cite{6} and the solutions are seen to exist for a particular range of the parameters only. In both the theories, we find that a whole series of BPs exists in the PDs of the theory. Numerically, we have determined up to $~4~$ BPs. 

In this work we concentrate on a theory with a massless complex scalar field in the absence of a cosmological constant.
The action of the theory with a massless scalar field is defined as:
\begin{eqnarray*}
S &= & \int \left[ \frac{R}{16\pi G} + {\cal L}_{M} \right] \sqrt{-g} ~  d^{4}~ x ~,~~ 
{\cal L}_{M} = \left[ - \frac{1}{4} F^{\mu\nu} F_{\mu\nu} 
- (D_{\mu} \Phi)^{\star} (D^{\mu} \Phi) - V(|\Phi|) \right] \\
D_{\mu} \Phi &=& (\partial_{\mu} \Phi + i e 
A_{\mu} \Phi) ~,~~ F_{\mu\nu} = (\partial_{\mu} 
A_{\nu} - \partial_{\nu} A_{\mu}) ~,~~
 V(|\Phi|) := \lambda |\Phi|
\end{eqnarray*} 
The asterisk in the above equation denotes complex conjugation. Here $~R~$ is the Ricci curvature scalar, $~G~$ is Newton's gravitational constant and $~g~$ = det($g_{\mu\nu}~)$, where $~g_{\mu\nu}~$ is the metric tensor. These studies not only confirm the existence of a sequence of BPs in the PDs but they also indicate that there is in fact an infinite series of such BPs (as is seen in our numerical results) \cite{5, 6}. The equations of motion (EoM) obtained via the variational principle are:
\begin{eqnarray*}
G_{\mu\nu} & \equiv & \bigg[R_{\mu\nu} - \frac{1}{2}
g_{\mu\nu} R\bigg] = (8 \pi G) T_{\mu\nu} \\
\partial_{\mu} \bigg( \sqrt{-g} F^{\mu\nu} \bigg) &=& -i ~ e \sqrt{-g} ~ \bigg[ \Phi^{\star} 
(D^{\nu} \Phi) - \Phi (D^{\nu} \Phi)^{\star}\bigg] \\
D_{\mu} \bigg( \sqrt{-g}  D^{\mu} \Phi \bigg) 
& = & \bigg[ + \frac{\lambda}{2} \sqrt{-g} ~ \frac{\Phi} {|\Phi|} \bigg] ~,~~
\bigg[ D_{\mu} \bigg( \sqrt{-g} D^{\mu} \Phi \bigg) \bigg]^{\star} =  + \bigg[ \frac{\lambda}{2} \sqrt{-g} ~ \frac{\Phi^{\star}}{|\Phi|} \bigg] 
\end{eqnarray*}
The energy-momentum tensor $~T_{\mu\nu}~$ is given by
\begin{eqnarray*}
T_{\mu\nu} &=& \bigg[ \bigg( F_{\mu\alpha} F_{\nu\beta}  g^{\alpha\beta} - \frac{1}{4} g_{\mu\nu} F_{\alpha\beta} F^{\alpha\beta} \bigg) +\bigg( (D_{\mu} \Phi)^{\star} (D_{\nu} \Phi) + (D_{\mu} \Phi) (D_{\nu} \Phi)^{\star} \bigg)\nonumber \\ 
& & ~~~~~~~~~~ - g_{\mu\nu} \bigg( (D_{\alpha} \Phi)^{\star} (D_{\beta} \Phi) \bigg) 
g^{\alpha\beta} - ~ g_{\mu\nu}~ V( |\Phi|) \bigg]
\end{eqnarray*}
To construct spherically symmetric solutions
we adopt a static spherically symmetric metric with Schwarzschild-like coordinates as:
$$ds^2= \bigg[ -A^2 N dt^2 + N^{-1} dr^2 +r^2(d\theta^2 + \sin^2 \theta \;d\phi^2) \bigg].$$
The components of the Einstein tensor $~G_{\mu\nu}~$ are obtained as: 
\begin{eqnarray*}
G_t^t &=& \biggl[ \frac{-\left[r\left(1-N\right)\right]'}{r^2} \biggr] , \, \,
G_r^r = \biggl[ \frac{2 r A' N -A\left[r\left(1-N\right)\right]'}{A\ r^2} \biggr],~~ 
G_\theta^\theta = \biggl[ \frac{2r\left[rA'\ N\right]' + \left[A\ r^2 N'\right]'}{2 A\ r^2} \biggr]
\  \   = \  G_\varphi^\varphi\,.
\end{eqnarray*}
Here the prime denotes differentiation with respect to $~r~$ and the arguments of $~A(r)~$ and $~N(r)~$ have been suppressed. Also, the Hilbert-Einstein equation expressed in component form reads: 
\begin{eqnarray*}
G_t^t &=& 8 \pi G\ T_t^t ~,~~~ G_r^r =  8 \pi G\  T_r^r \ \, ~,~~~ 
G_\theta^\theta =  8 \pi G \,T_\theta^\theta \ ,\  ~,~~~
G_\varphi^\varphi =  8 \pi G\ T_\varphi^\varphi ~.
\end{eqnarray*}
We now work under the assumption of a vanishing magnetic field 
i.e., we assume that ($~{\overrightarrow{B}}(x^\mu) = \overrightarrow{\nabla} \times {\overrightarrow{A}} (x^\mu) = 0~$). Then only the time-like component of 
$~A_{\mu}~$ is non-vanishing. For the matter fields we make the following Ans\"atze: 
$ \Phi(x^\mu)=\phi(r) e^{i\omega t} ~,~~
A_\mu(x^\mu) dx^\mu = A_t(r) dt. ~$
The components of the energy-momentum tensor could now be evaluated using the above Ans\"atze. 

Next, we introduce the dimensionless constant parameters and dimensionless field variables with dimensionless arguments through the following re-definitions (where $~a~$ is dimensionless):
\begin{equation*}
 \beta=\frac{\lambda\,e}{\sqrt{2}}   \ \ \ ,\ \ \ \alpha^2 (~:=a) = \frac{4\pi G\,\beta^{2/3}}{e^2} 
\end{equation*} 
Also we redefine the functions $~\phi(r)~$ and $~A_t(r)~$ through the following relations:
\begin{equation*}
 h(r)=\frac{(\sqrt{2} \;e\, \phi(r))}{\beta^{1/3}} \ \ \ ,\ \ \  b(r)=\frac{(\omega+e A_t(r))}{\beta^{1/3}} . \label{hb}
 \end{equation*} 

The metric functions $~A(r)~$ and $~N(r)~$ here are already dimensionless. We also introduce a dimensionless coordinate 
$~\hat{r}~$ defined by $~\hat{r}:=\beta^{1/3}\,{r}~$ (which in turn implies $~\frac{d}{d{r}}=\beta^{1/3}\frac{d}{d\hat{r}}~$). Above equations in terms of dimensionless coordinate $~\hat{r}~$ become: 
\begin{equation*}
 h(\hat{r})=\frac{(\sqrt{2} \;e\, \phi(\hat{r}))}{\beta^{1/3}} \ \ \ ,\ \ \  b(\hat{r})=\frac{(\omega+e A_t(\hat{r}))}{\beta^{1/3}} . \label{hb1} 
 \end{equation*} 

From now onwards, we change the arguments of the field variables from the dimension-full radial coordinate $~r~$ to the dimensionless radial coordinate $~\hat{r}~$.
The matter field equations of motion involving the dimensionless field variables as well as their derivatives (with $~{\rm sign }(h)~$ denoting the usual signature function) then read:
\begin{eqnarray*}
\left[A N \hat{r}^2 h'\right]' = \frac{\hat{r}^2}{AN}\left(A^2N {\rm sign}(h) -b^2 h\right) ~,~~
\left[\frac{\hat{r}^2 b'}{A}\right]' = \frac{b h^2 \hat{r}^2}{AN}. 
\end{eqnarray*} 
Primes here denote differentiation with respect to dimensionless radial coordinate $~\hat{r}~$. 
Eventually, the set of EoM (expressed in dimensionless constant parameters and dimensionless field variables and their derivatives involving dimensionless arguments) which is to be solved numerically is:
\begin{eqnarray}
h'' & = &\bigg[ \frac{\alpha^2\, \hat{r} h'}{A^2N} \left(2 A^2 h + \,b'^2\right)
-\frac{h'\big(1 + N \big)}{\hat{r}N}
+ \frac{A^2 N \, {\rm sign}(h) - b^2 h}{A^2 N^2}\bigg]\label{eq_H}\nonumber\\
b'' & = & \bigg[\frac{\alpha^2}{A^2 N^2} \hat{r} b'\left(A^2 N^2 h'^2 + b^2 h^2\right) -\frac{2 b'}{\hat{r}} + \frac{b h^2}{N}\bigg]\label{eq_b}\nonumber\\
N' & = &\bigg[ \frac{1-N}{\hat{r}} -\frac{\alpha^2 \hat{r}}{A^2 N}
\bigg(A^2 N^2 h'^2 + N b'^2 +  b^2 h^2
+ 2 A^2 N h \bigg)\bigg] \label{eq_N}\nonumber\\
A'  &=& \bigg[ \frac{\alpha^{2}\hat{r}}{A N^2}\left(A^2 N^2 h'^2 + b^2 h^2 \right) \bigg]\ \label{eq_A}\nonumber
\end{eqnarray}
The above Eqs. are solved numerically by introducing a new coordinate $~x~$ as follows:
$$~\boxed{~~~\hat{r}~~=~~\hat{r}_{i} ~+~~ x ~(\hat{r}_{o}-\hat{r}_{i})\ , \ \ \ \ 0\leq x \leq 1 ~~~} ~$$
where $~\hat{r}_{i}~$ and $~\hat{r}_{o}~$ are the inner and outer radii of the shell. This  implies that $~\hat{r}=\hat{r}_{i} \ \ {\rm at} \ \ x = 0~$ and $~\hat{r} = \hat{r}_{o}\  {\rm at} \ x=1~$.
Thus inner and outer boundaries of the shell are always at $~x=0~$ and $~x=1~$ respectively, while their radii $~\hat{r}_{i}~$ and $~\hat{r}_{o}~$ become free parameters. 

Also, for boson stars $~\hat{r}_{i} = 0. ~$ We then solve these equations under specific boundary conditions (BCs): $~A(\hat{r}_{\rm o}) = 1 ~,~~ N(0) = 1 ~,~~ b'(0) = 0 ~,~~ h'(0) = 0 ~,~~ h(r_{\rm o}) = 0 ~,~~ h'(\hat{r}_{\rm o})= 0.~$ For the boson stars we match the exterior region $~\hat{r} > \hat{r}_o ~$, with the Reissner-Nordstr\"om solutions. For the boson shell solutions (with empty space-time in the interior of the shells), with $~\hat{r} < \hat{r}_{i}~$, we choose the BCs as: 
$~A(\hat{r}_{\rm o}) = 1 ~,~~ N({\hat{r}}_{i}) = 1 \,, \ \ \ b'(\hat{r}_{i})=0 \ ,\ \  h(\hat{r}_{i})= 0 ~,~~ h'({\hat{r}}_{i}) = 0 ~,~~ h(\hat{r}_{o}) = 0 ~,~~ h'(\hat{r}_{o})= 0 , ~$ where $~\hat{r}_{i}~ $ and $~\hat{r}_{o}~$ are the inner and outer radii of the shell. Here, we match the interior region 
$~\hat{r} < \hat{r}_{i} $~ and the exterior region $~\hat r > \hat r_{o}~$, with the appropriate Reissner-Nordstr\"om solutions.
The U(1) invariance of the  theory leads to a conserved Noether current:
\begin{eqnarray*}
j^{\mu} = - i \,e\,\left[\Phi(D^{\mu} \Phi)^{\star} - \Phi^{\star} (D^{\mu} \Phi) \right]\ ,\   \
j^{\mu}_{\ ;\mu} = 0
\end{eqnarray*} 
Its time component corresponds to charge density. The global charge $~Q~$ of the boson star is given by: 
\begin{equation*}
Q = - \frac{1}{4\pi}\int_{0}^{\hat{r}_o} j^t \sqrt{-g} \,dr\,d\theta\,d\phi  \,,\  
j^t = - \frac{h^2(\hat{r}) b(\hat{r})}{A^2(\hat{r}) N(\hat{r})} .\label{charge} 
\end{equation*} 
One can read off the mass from the metric as usual, making use of the fact, that the metric outside the compact star corresponds to a Reissner-Nordstr\"om metric. The mass $~M~$ of the boson star solutions is given by: 
$~M = \biggl(1-N(\hat{r}_o)+\frac{\alpha^{2} Q^{2}}{\hat{r}_o^2}\biggr)\frac{\hat{r}_o}{2}.~$

As explained in Ref.~\cite{9}, an extremal Reissner-Nordstr\"om metric would satisfy a proportionality between the mass $~M$~ and the charge $~Q~$ given by $~M = \alpha Q = \sqrt{a} Q~$, where $~ \alpha (:= \sqrt{a}) ~$ enters here because of the units employed (whereas in the usual geometric units the extremal solution satisfies $~M = Q~$). We can now consider three different cases for the exterior solution ($r~>~r_o$) outside the bosonic matter distribution: (i) case $~M/Q ~<~ \sqrt{a}~$ corresponds to a horizonless (naked) Reissner-Nordstr\"om solution (we note, that naked solutions would also arise, when the mass and charge of known elementary particles would be inserted on the left hand side of this relation). (ii) Case $~M/Q = \sqrt{a}~$ corresponds to an extremal Reissner-Nordstr\"om solution (we note, that the sets of boson star solutions end, when such an extremal Reissner-Nordstr\"om solution is reached and $b(0)=0$).
(iii) Case $~M/Q ~>~ \sqrt{a}~$ corresponds to a solution where the radius of the boson star is greater than its putative event horizon radius $~{\hat{r}_H}~$ (implying that the boson star is well outside the range of becoming a black hole. This discussion is reflected in the results shown in  Fig. $7$, which shows plots of various fields with respect to $~\hat{r},~$ 
implying thereby that the boson stars either have a radius $~\hat{r}~ >~ {\hat{r}_H}~$, or that they are smoothly matched to a naked Reissner-Nordstr\"om solution for $~M/Q < \sqrt{a}~$. The case $~M/Q = \sqrt{a}~$ is not exhibited in Fig. $7$. It would correspond to a solution where a throat develops and $b(0)=0$ (cf. Ref. \cite{9}).  

For a study of ~QCD~ boson stars \cite{10}, we include the QCD Lagrangian density in our theory:
$~{\cal L}_{QCD} = ( - \frac{1}{4} \tilde{\mathcal{F}}_{\mu\nu}^{a} \tilde{\mathcal{F}}^{\mu\nu}_{a}~) + $ (a fermionic piece), where $~ \tilde{\mathcal{F}}_{\mu\nu}^{a} = [(\partial_{\mu} A_{\nu}^a - \partial_{\nu}A_{\mu}^{a}) + g f^{abc} A_{\nu}^b A_{\nu}^c].~$ Here $~A_{\mu}^{a}~$ are the gluon fields and $~ a = 1, 2, 3,...., 8~~$ for the $~8~$ gluonic degrees of freedom (all symbols have their usual meanings). The fermionic piece of the Lagrangian density however, vanishes identically in the case of boson stars (because boson stars do not have any fermions) leaving us only with the gluonic part: $~ {\cal L}_{QCD} =  - \frac{1}{4}  \tilde{\mathcal{F}}_{\mu\nu}^{a} \tilde{\mathcal{F}}^{\mu\nu}_{a} ~.~$  We now assume a vanishing chromo-magnetic field which implies: 
$~\overrightarrow{B}^{a} = \overrightarrow{\nabla} \times \overrightarrow{A}^{a} = 0.~$ This leaves us only with the time-like component of $~A_{\mu}^{a}~$ (which is non-vanishing) and the space-like components of  $~A_{\mu}^{a}~ $ do not contribute anything under this assumption. The energy-momentum tensor $~{T}_{\mu\nu}~$ changes under this assumption and one also obtains an additional EoM for $~A_{\mu}^{a}~.~$
The phase space of the theory now gets enlarged because of the inclusion of the QCD Lagrangian. The field variables of this new theory comprise the metric, the scalar fields, the electromagnetic field and the gluon fields. Such studies require a rather careful investigation and its details would be published later.

Our numerical results are shown in our various plots \cite{5, 6} for different values of the parameter $~a~$ in the range $~a = 0.0~$ to $~a = 0.2500.~$. The insets in all our figures magnify particular interesting areas of the figures and the astrisks represent the transition from the boson stars to boson shells. For further details of our phase diagrams and the BPs, we refer to the work of our Refs.~\cite{5, 6}. We obtain $~4~$ BPs in the phase diagrams of the theory \cite{9} as shown in Fig. $1$ and Fig. $2$. The results in Fig. $1$ show the first three BPs and the $4$th BPs is shown in Fig. $2$. Numerical calculations become increasingly challenging as one progresses from one BP to the next BP (even for getting the $~4$th BP we have to go to the 8th digit of numerical accuracy). The asterisks in the above figures reside on the axis $~b(0)~$, corresponding to $~h(0)=0~$, and they mark the transition points from the boson stars to boson shells. 
We present our results on the study of (i) the phase diagrams of the theory showing BPs in  Fig. $1$ and Fig. $2$ (ii) variations of $~M~$ and $~Q~$ with $~\hat{r}_{0} ~$ in Fig. $3$ and Fig. $4$ (iii) variations of $~M/Q~$ with $~Q~$ in Fig. $5$ and Fig. $6$ and (iv) distribution of various fields with $~\hat{r}_{0}~$ (for different values of the dimensionless variable $~a~$ in the range $~a = 0.050~$ to $~a = 0.250~$) in Fig. $7$.

Based on the regularity of the occurence of these BPs we conjecture \cite{9} that perhaps a whole sequence of BPs may be waiting to be discovered (which might display a self-similar pattern). In the set of BPs $~a_{c_1} ,a_{c_2}~$ and $~a_{c_3}~$, one notices that the difference between the BPs  $~\triangle_{n} =  (a_{c_n} - a_{c_{n-1}})~~$ decreases strongly and an exponential approach towards a limiting value $~a_{c_\infty}~$ could be expected. We thus propose an Ansatz for the $~n$th BP and it seems to work \cite{9}:
\begin{equation*}
\boxed { ~~~ a_{c_n} ~= ~~~ a_{c_\infty} ~ + ~f~exp(- ~b~n) ~~~~ = ~~~~ a_{c_\infty} ~+~f~a^{n} ~~~}
\end{equation*}
Where $~a_{c_\infty} ~,~~ f~$ and $~ b~$ (or $~a ~=~ exp(-b)$) are constants. We thus conclude that indeed an infinite sequence of BPs exists. Using the first 3 BPs and making a prediction for the $~4$th BP, we indeed find the $~4$th BP to be located at $~ a_{c_4} = 0.16827861,~~$ which agrees with its predicted value. 

Fig. $7$ shows the distribution of field variables
 $~A(\hat{r})~$, $~h(\hat{r})~$, $~b(\hat{r})~$ and $~N(\hat{r})~$ with respect to the dimensionless radial coordinate $~\hat{r}$, which are solutions of the final set of four coupled nonlinear differential equations. 
The electromagnetic field represented by $~b(\hat{r})~$ and the gravitational field represented by the metric functions $~A(\hat{r})~$ and $~N(\hat{r})~$ correspond to long range forces. 
Fig. $8$ shows the exponential fit (solid line) for the $~4$ BPs (asterisks) obtained from PDs of the theory. The difference of the predicted values and the values obtained from PDs are seen to be negligible.
\begin{figure}[!tbp]
\centering
\begin{minipage}[b]{0.4\textwidth}
\includegraphics[width=\textwidth]{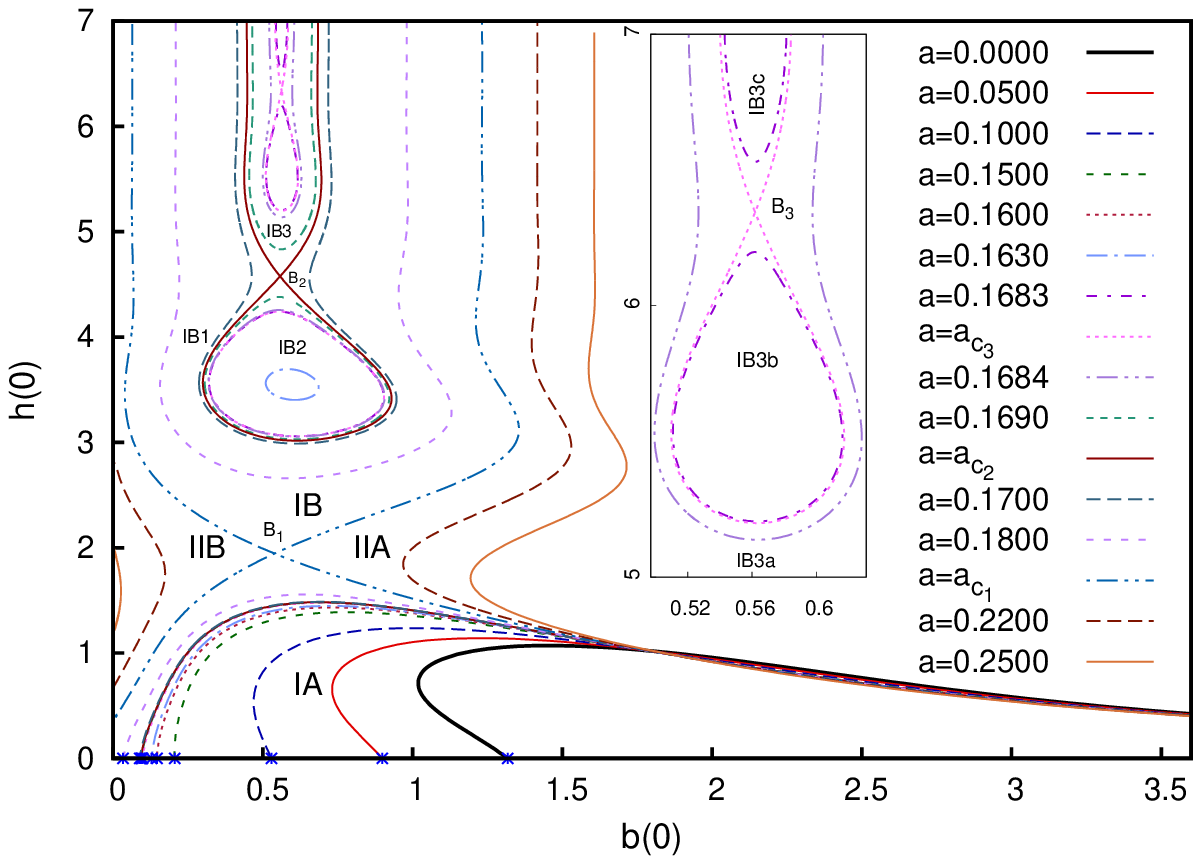}
\caption{
2D phase diagram of the theory showing $3$ BP's in the plot of $~h(0)~$ versus $~b(0)~$ for different values of $~a~$.}
\end{minipage}
\hspace{1.0cm}
\begin{minipage}[b]{0.4\textwidth}
\includegraphics[width=\textwidth]{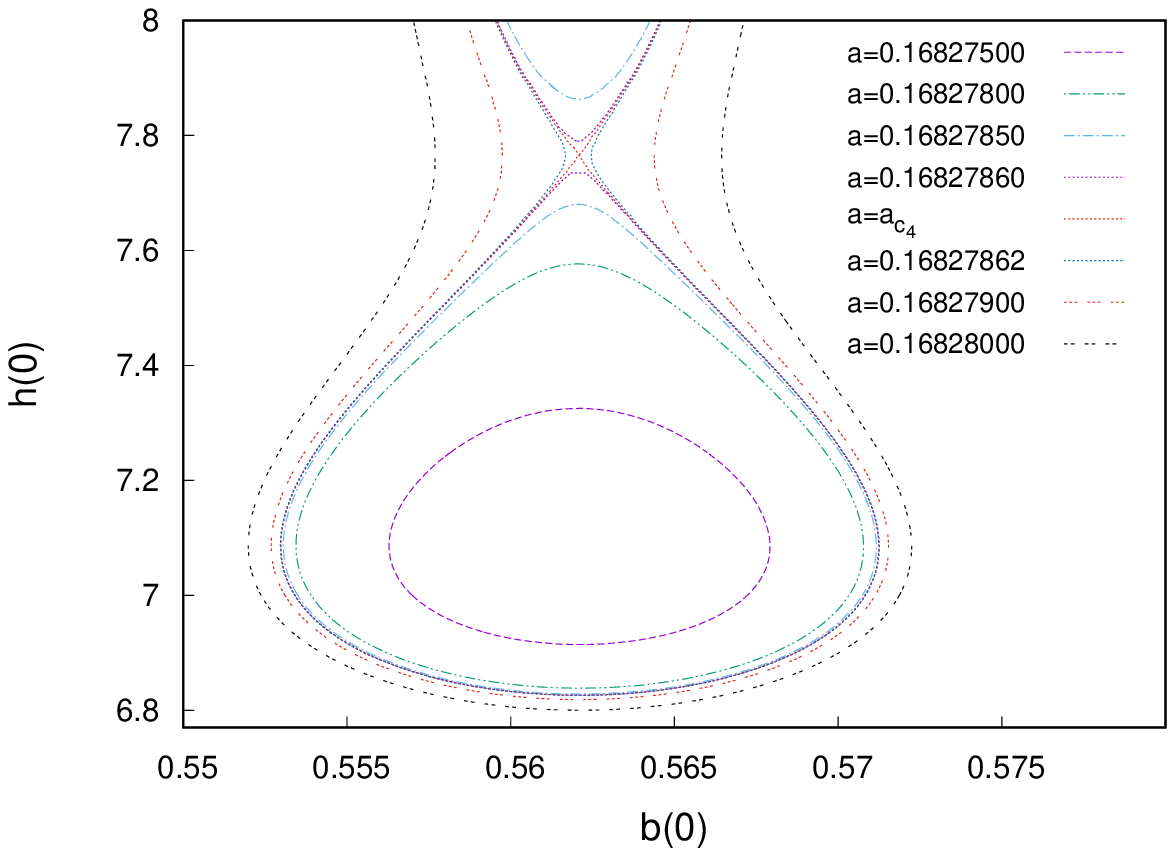}
\caption{
2D phase diagram of the theory showing $4$th BP in the plot of $~h(0)~$ versus $~b(0)~$ for different 
values of $~a~$.}
\end{minipage}
\end{figure}
\begin{figure}[!tbp]
\centering
\begin{minipage}[b]{0.4\textwidth}
\includegraphics[width=\textwidth]{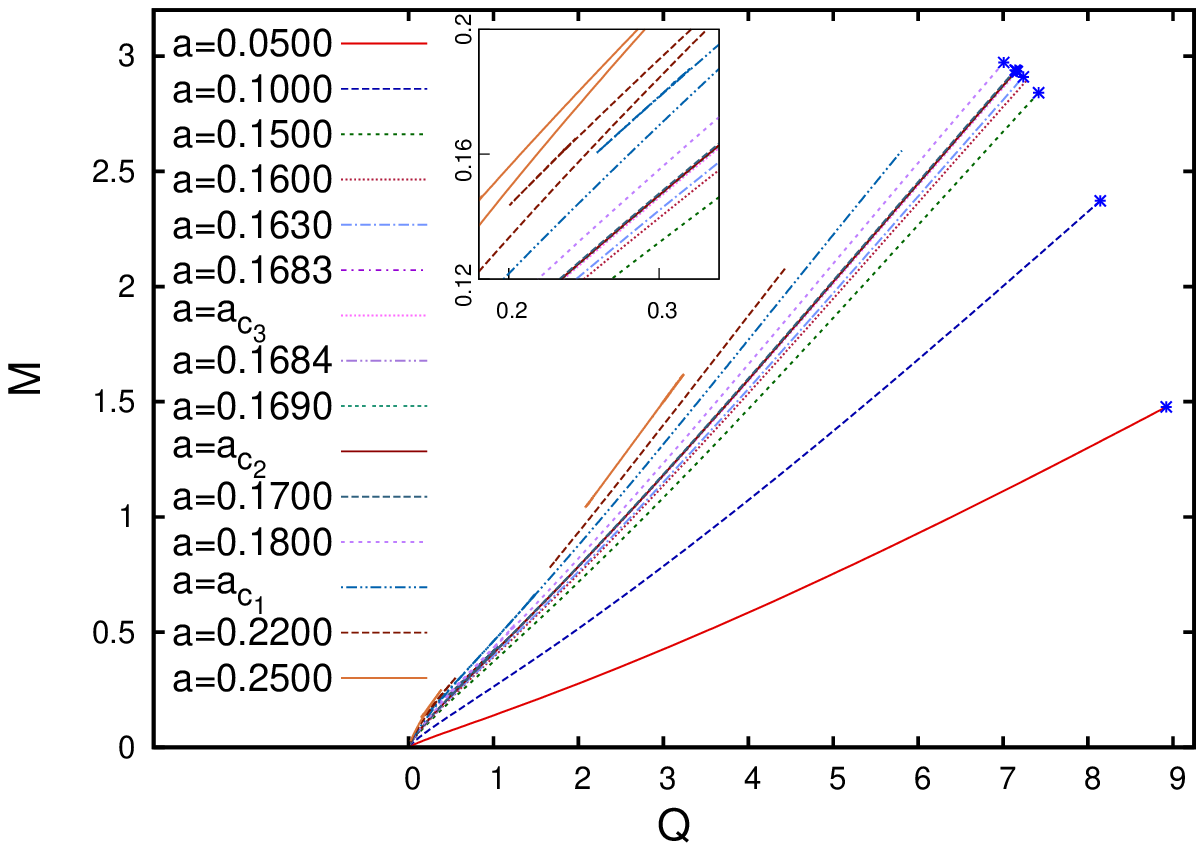}
\caption{
Plot of $~M~$ versus $~Q$ for different values of $~a~$ }
\end{minipage}
\hspace{1.0cm}
\begin{minipage}[b]{0.4\textwidth}
\includegraphics[width=\textwidth]{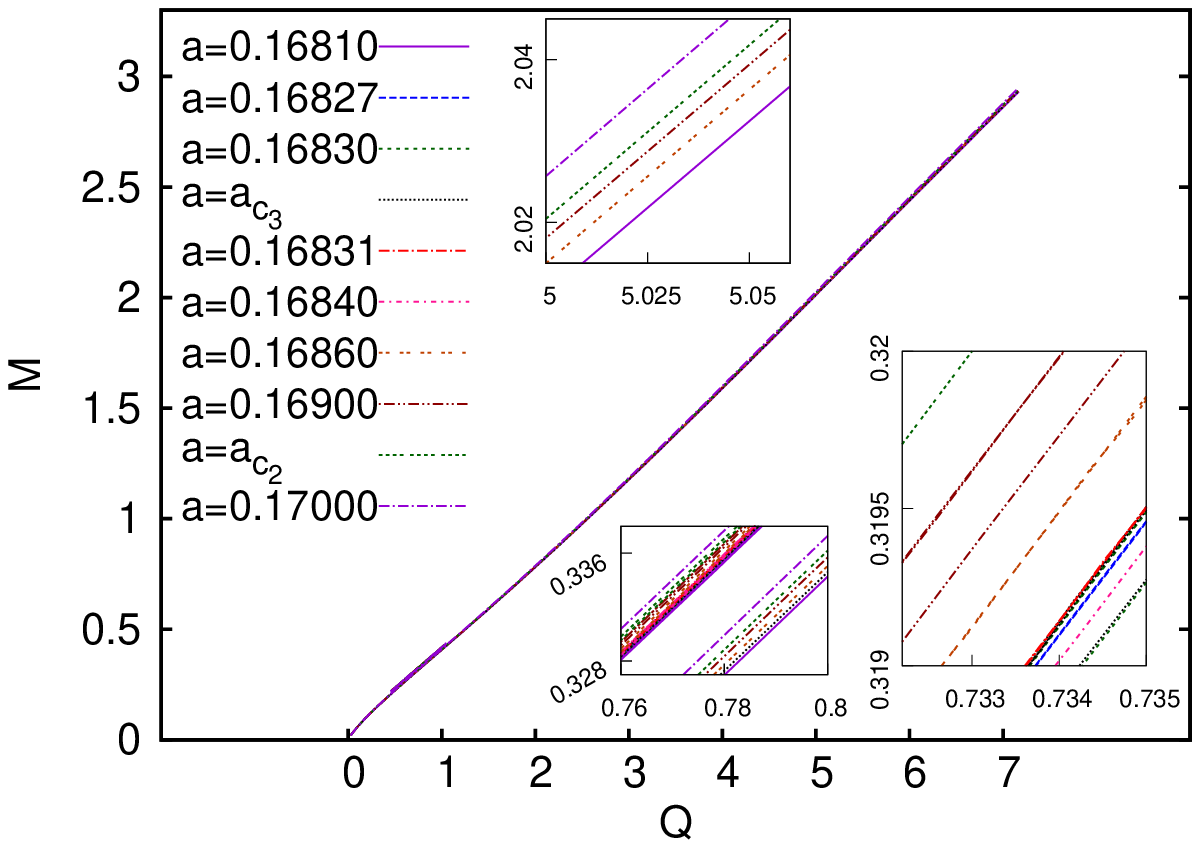}
\caption{
Plot of $~M~$ versus $~Q~$ for different values of $~a~$}
\end{minipage}
\end{figure}
\begin{figure}[!tbp]
\centering
\begin{minipage}[b]{0.4\textwidth}
\includegraphics[width=\textwidth]{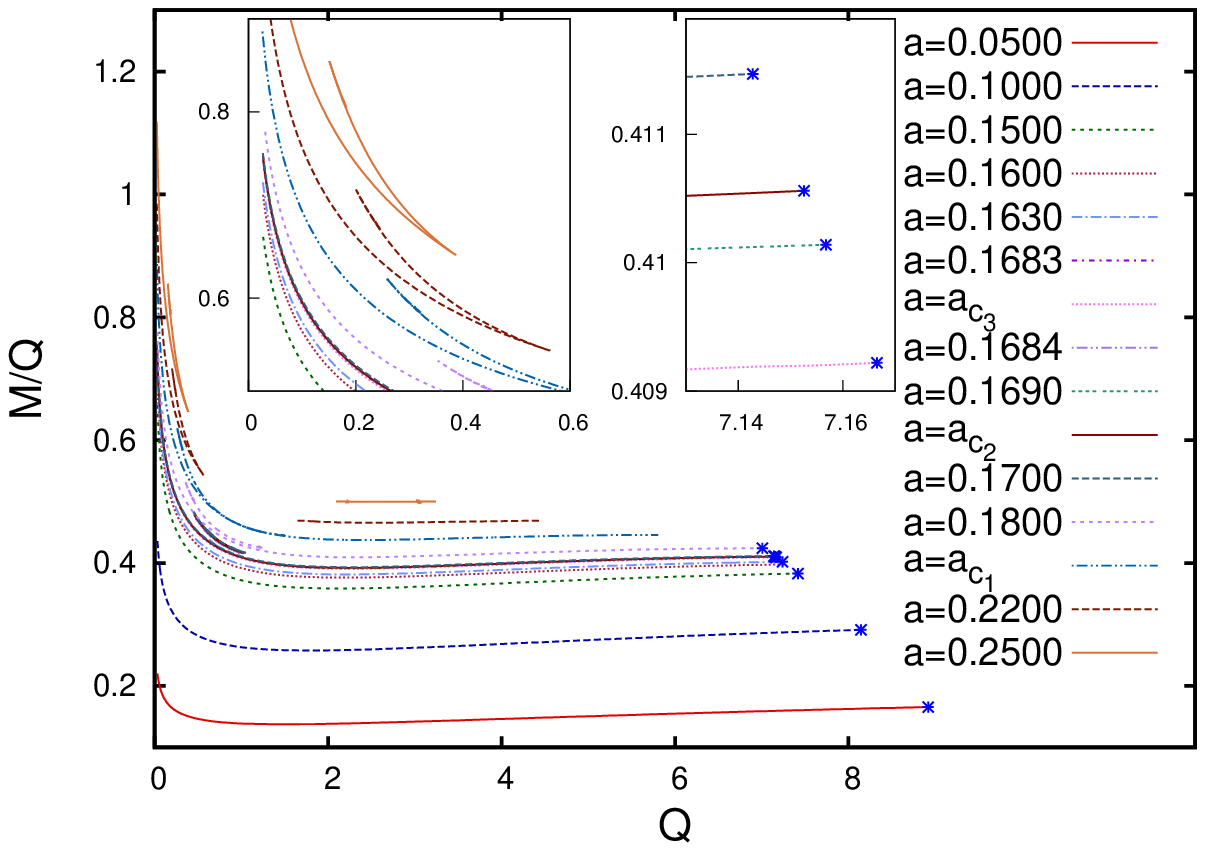}
\caption{
Plot of $~M/Q~$ versus $~Q~$ for different values of $~a~$}
\end{minipage}
\hspace{1.0cm}
\begin{minipage}[b]{0.4\textwidth}
\includegraphics[width=\textwidth]{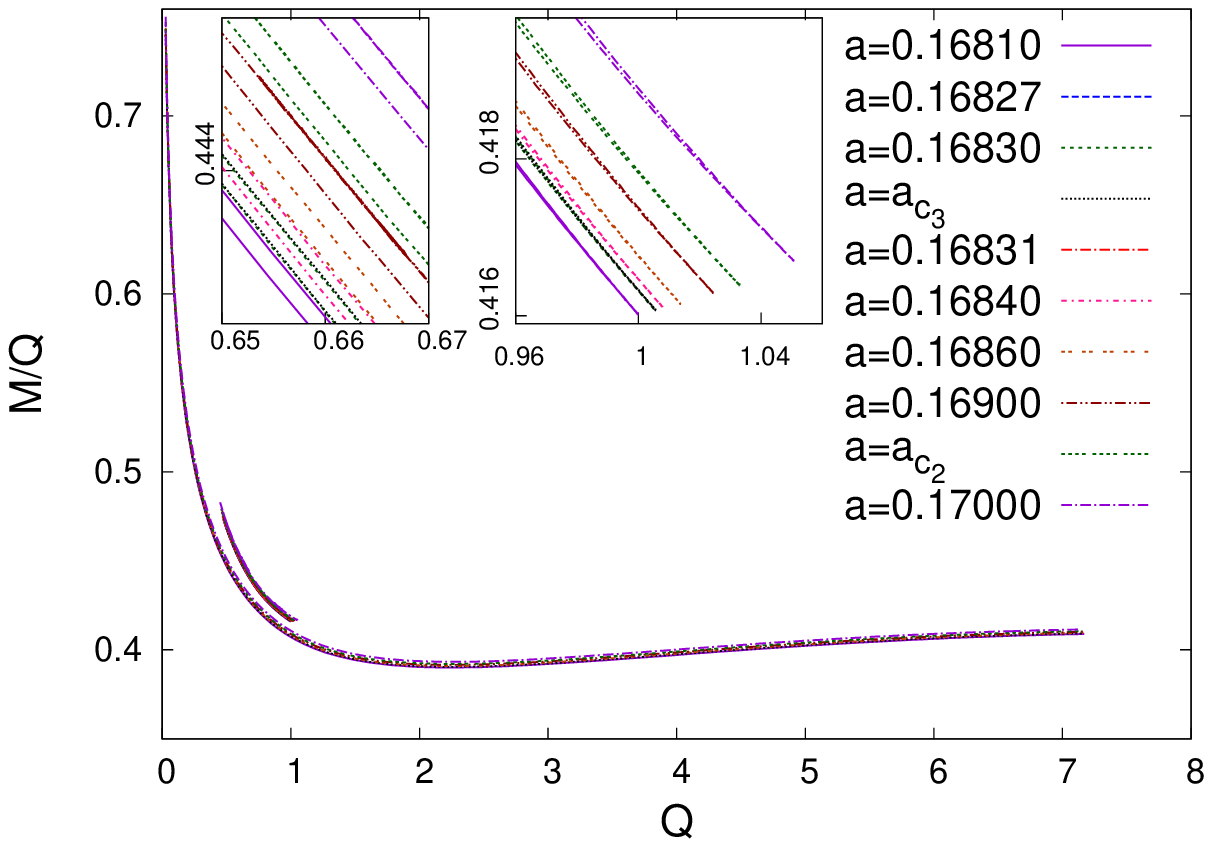}
\caption{
The plot of $~M/Q~$ versus $~Q~$   for different values of $~a~$}
\end{minipage}
\end{figure}
\begin{figure}[!tbp]
\centering
\begin{minipage}[b]{0.4\textwidth}
\includegraphics[width=\textwidth]{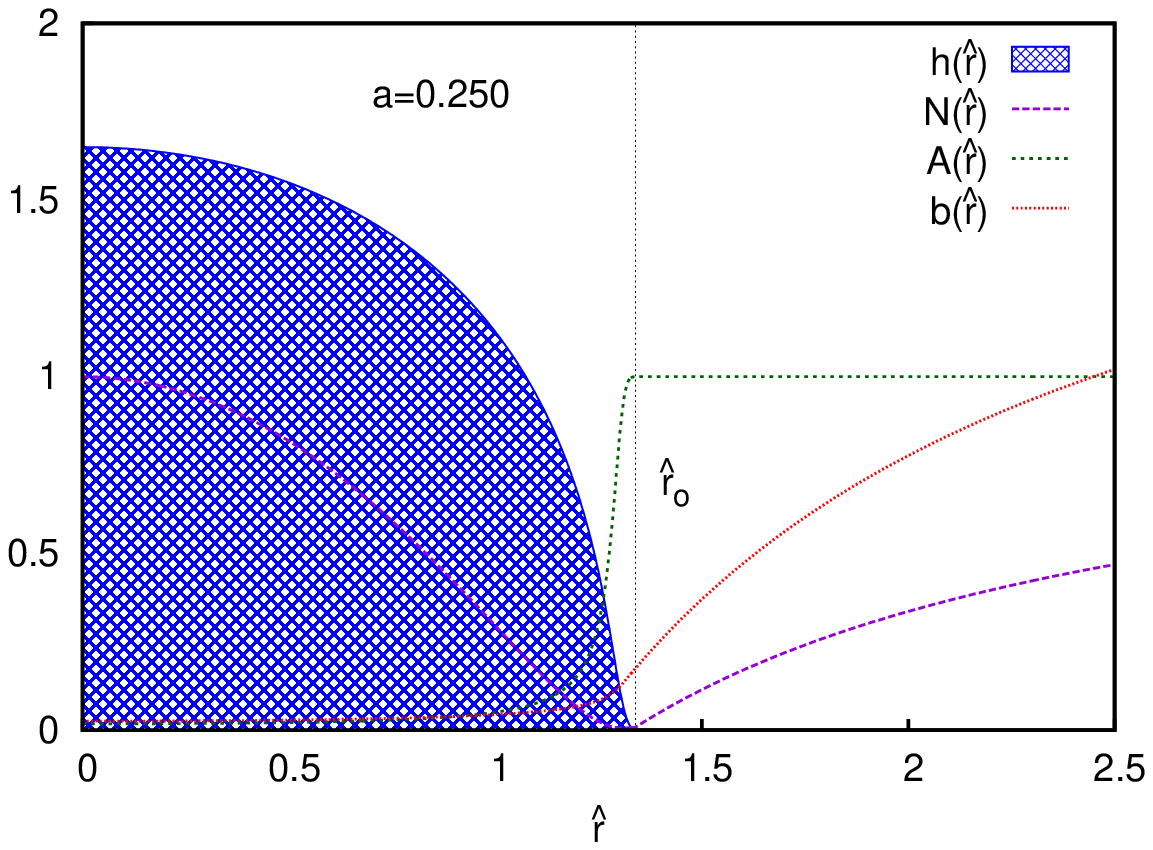}
\caption{
Distribution of four fields $A(\hat{r})$, $h(\hat{r})$, $b(\hat{r})$ and $N(\hat{r})$) with respect to the dimensionless radial coordinate 
$~\hat{r}~,$~ for $~a~ = 0.250.$}
\end{minipage}
\hspace{1.0cm}
\begin{minipage}[b]{0.5\textwidth}
\includegraphics[width=\textwidth]{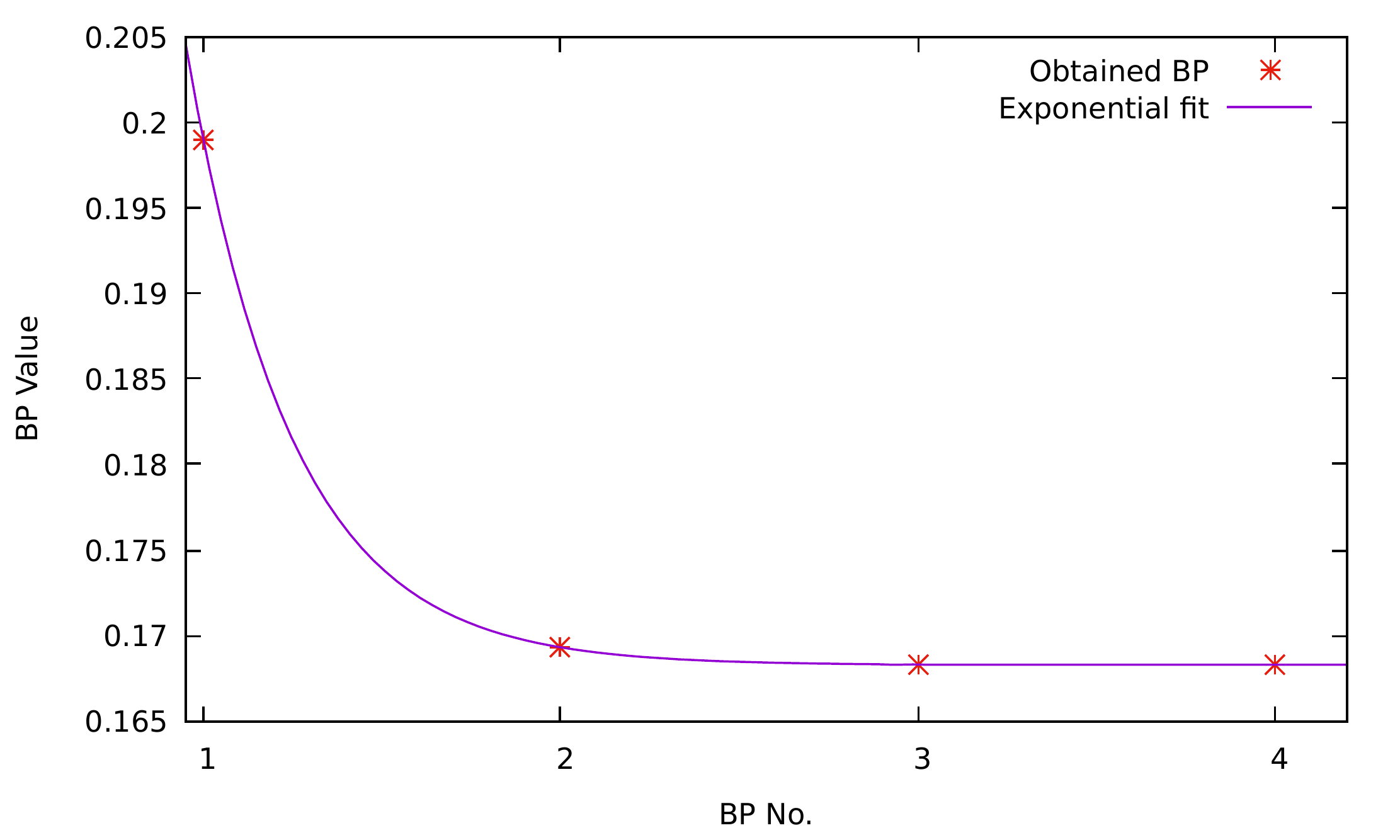}
\caption{
Exponential fit (solid line) for the $4$ bifurcation points (asterisks) obtained from the phase diagrams of the theory.}
\end{minipage}
\end{figure}

We thank Cedric Lorce and Arkadius Trawanski, the organizers of the International Conference on Light Cone Physics (LC-2019) held at Ecole Polytechnique Palaiseau France, during September 16-20, 2019, for their great hospitality.

\end{document}